# Behavior of molecules and molecular ions near a field emitter

Baptiste Gault[1,2,*], David W. Saxey[3], Michael V. Ashton[4], Susan B. Sinnott[5], Ann N. Chiaramonti[6], Michael P. Moody[1], Daniel K. Schreiber[7]

[1] *Department of Materials, University of Oxford, Parks Road, Oxford, OX13PH, UK*

[2] *Max-Planck-Institut für Eisenforschung, Department of Microstructure Physics and Alloy Design, 40237 Düsseldorf, Germany*

[3] *John de Laeter Centre, Curtin University, Perth, Western Australia 6102, Australia*

[4] *Department of Materials science and Engineering, University of Florida, Gainesville, FL 32611-6400, USA*

[5] *Materials Science and Engineering, Pennsylvania State University, State College, PA 16802, USA*

[6] *Material Measurement Laboratory, National Institute of Standards and Technology, Boulder, Colorado 80305, USA*

[7] *Energy and Environment Directorate, Pacific Northwest National Laboratory, Richland, WA 99352, USA*

* Corresponding author: b.gault@mpie.de; This work is a partial contribution of the U.S. Government and therefore is not subject to copyright in the United States.

# ABSTRACT

The cold emission of particles from surfaces under intense electric fields is a process which underpins a variety of applications including atom probe tomography (APT), an analytical microscopy technique with near-atomic spatial resolution. Increasingly relying on fast laser pulsing to trigger the emission, APT experiments often incorporate the detection of molecular ions emitted from the specimen, in particular from covalently or ionically bonded materials. Notably, it has been proposed that neutral molecules can also be emitted during this process. However, this remains a contentious issue. To investigate the validity of this hypothesis, a careful review of the literature is combined with the development of new methods to treat experimental APT data, the modelling of ion trajectories, and the application of density-functional theory (DFT) simulations to derive molecular ion energetics. It is shown that the direct thermal emission of neutral molecules is extremely unlikely. However, neutrals can still be formed in the course of an APT experiment by dissociation of metastable molecular ions.

## I. INTRODUCTION

The emission of particles under intense electric fields is a well-documented concept, which was first explained soon after the establishment of quantum mechanics. In 1928 Oppenheimer, Fowler-Nordheim, Gamow, and Gurney and Condon [1–4], independently, described the emission of electrons or positively charged particles ($\alpha$, protons, etc.). Subsequently, Erwin W. Müller showed that an intense standing electric field, in the range of $10^{10}$ V m$^{-1}$ to $6\times10^{10}$ V m$^{-1}$, facilitates the field-induced removal of metal surface atoms from a substrate of the same element [5]. This process is referred to as field evaporation. The intense electric fields required for this process can be generated at the tip of a sharp needle-shaped specimen (20 nm to 200 nm in diameter) subjected to a high voltage (2 kV to 15 kV).. This underpins the materials characterization technique called the atom probe [6]. In atom probe experiments, time-control of the field evaporation process is achieved by using high-voltage[6] or laser pulses [7–9] superimposed on a static standing field, as depicted in Figure 1~~Figure 1~~. This enables the elemental identification of each evaporated ion by time-of-flight mass spectrometry. Building upon this work, Cerezo et al. [10], followed by Blavette et al. [11], and finally Kelly et al. [12] developed a technique currently known as atom probe tomography (APT). APT provides analytical, three-dimensional mapping of a range of solid materials [11,13–25] with near-atomic resolution.

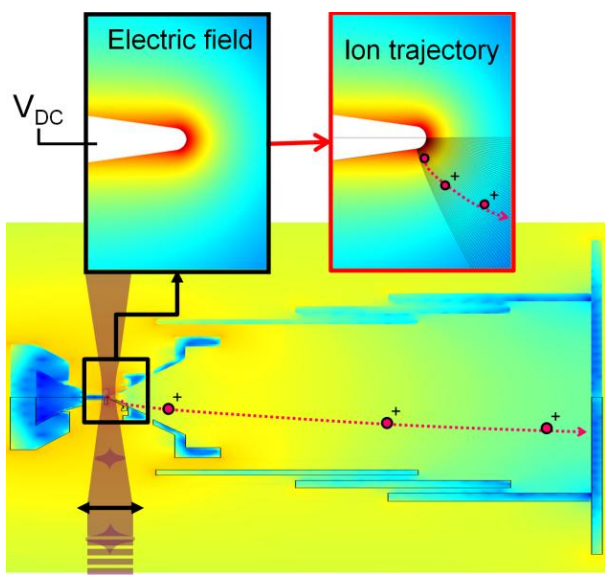

*Figure 1: Schematic view of an atom probe microscope, with the ion source subjected to a high voltage and illuminated by laser pulses, triggering the field evaporation of ions that strike a single-particle detector. The colour-scale corresponds to the distribution of the potential within the chamber as calculated by boundary-element methods. For more details, see ref.[26].*

In the APT analysis of metals, field evaporation in vacuum leads to the emission of singly-charged (or in some cases, doubly-charged) mono-atomic ions which can then potentially be post-ionized once or even multiple times [27–29]. In contrast, in the case of field evaporation of covalently or ionically bonded materials, charged molecules (molecular ions) are regularly detected [30–32]. As detailed in thorough review articles by Mathur [33,34], molecular ions have been the subject of intense studies in the field of mass spectrometry as they can form due to the impact of other charged particles or photons. Molecular ions are metastable and usually dissociate into smaller fragments. Beyond its fundamental interest, the dissociation of molecular ions is also a commonly encountered problem in mass spectrometry of (e.g.) organic compounds, and computational methods have often been used to interpret experimental observations [35]. In addition, molecular ions are known to be very reactive [36] and the dissociative recombination [37] of molecular ions is known to play a role in atmospheric and spatial chemical processes [38].

The rate of publication of pulsed-laser APT data has surged in recent years[39,40]. In stark contrast, progress in understanding the fundamental physics of laser-assisted field evaporation, particularly for non-metals, has failed to keep pace. For metals, it is accepted that field evaporation is caused by a sharp increase in the specimen temperature due to the absorption of the light from each laser pulse, which is then quenched as the heat is transported inwards and then along the length of the shank [41] in a transient process that is often referred to as a thermal pulse [42,43]. The least conservative estimates for the case of a metal specimen predict temperature increases of up to a maximum of 600 K in tungsten [41] when the standing electric field is 75 % of the intensity required to field evaporate the specimen without laser illumination. However, these are conditions that are not expected to yield good APT performance [44] and are usually avoided.

Recent research indicated that the field evaporation mechanisms for semiconductors may differ to that of metals with high and fast phonon excitation that could result in very high temperature reached by the surface over very short durations [45,46] Importantly, building on previous work [47,48], Silaeva and co-workers [49] also proposed that the electric field causes semiconductor and insulator surfaces to actually behave like metallic surfaces, with some experimental evidence based on field ion imaging of insulating materials. From a field ionization perspective, it should therefore be possible to extend the interpretation of field evaporation processes observed in metals to these materials.

In field evaporation theory, little attention has been paid to the ionisation and dissociation behaviour of emitted molecules and molecular ions under high electric field, despite their growing influence on the analytical performance of APT [32,50]. In the context of APT, the lifetime of molecular ions can be sufficiently long to allow them to reach the detector. Indeed, molecular ions are commonly observed in APT analyses of a variety of materials, including even metals evaporated at very high laser intensities and low electric field [32,51–53]. In addition, under such conditions, it was proposed recently that neutral molecules in gaseous form, namely, $N_2$ and $O_2$, are emitted directly from the surface of the field emitter [54–60].

This study does not focus on the fundamentals of the field evaporation process, but rather it aims to investigate what happens to molecules or molecular ions that have been emitted during an APT experiment. First, building upon a careful and critical survey of the literature, it is shown that neutrals desorbing from the surface are very likely to be ionized under experimental conditions normally used for APT. Further, new analyses of experimental data are combined with simulations of ion trajectories to demonstrate that the dissociation of molecular ions can lead to the formation of neutrals. Finally, the energetics of complex molecular ions is explored. In particular, their stability and propensity for ionization are investigated. It is demonstrated that the concept of neutral formation by molecular dissociation can be generalised to a broad range of material analyses.

## II. NEUTRALS IN HIGH-ELECTRIC FIELDS: THE CASE OF GAN

The literature providing any direct evidence for the emission of neutrals in high electric field conditions, such as those necessary for field evaporation in the conditions of atom-probe operation, is scarce. Even studies were experimental results are obtained in conditions most likely to lead to the emission of neutrals, namely very low electric field conditions and therefore high maximum surface temperature, such as ref. [41,43,51], only report the detection of ions. Yet, several studies have proposed that the discrepancy between the APT measured composition of compound semiconductors and oxides and their actual stoichiometry was due to thermal desorption of gaseous molecules, $N_2$ and $O_2$, directly from the surface of the specimen [54–60]. Here we review the case of GaN, where such behaviour is a well-documented issue. GaN has been the focus of numerous APT studies in recent years due to its technological importance, and its well known, almost perfectly-defined stoichiometry with a Ga:N ratio of 1:1.

### A. COMPOSITIONAL ANALYSIS

Three recent articles report APT measurement of composition of N in GaN as a function of the laser pulse energy and/or standing voltage [61–63]. These results, as well as some additional newly reported data, are plotted in Figure 2Figure 2 and the detail of the instruments on which those were obtained can be found in the methods section at the end of this article as well as in the caption of the figure. Direct comparison between these studies is difficult, since different or insufficient parameters defining the experiment have been reported. This precludes comparing and interpreting the results in depth. For example, the laser pulse energy is reported rather than the intensity; voltages are specific to individual specimens; and various experimental protocols have been used, either varying the pulsed laser energy and the standing voltage so as to maintain a constant rate of detection, or keeping the laser pulse energy constant while increasing the voltage. Despite these complications, some common trends can be derived. Both a higher than expected amount of N and a higher level of background are observed at low laser power (high electric field). In contrast, higher than expected levels of Ga and lower background are both observed at high laser power (low electric field). It has been proposed that the unexpected behavior at low electric fields is caused by formation of a significant amount of neutral $N_2$ gas molecules at the surface, which are then desorbed or sublimated without ionization (i.e., remain as neutral $N_2$ molecules) and are hence never detected. With increasing electric field, the amount of dinitrogen ionized, and thus detected, progressively increases. However, in the case that the standing electric field is increased too much, Ga is field evaporated at the standing temperature uncorrelated to the laser pulse, which precludes its identification. These explanations are consistent with the reported measurements. However, the above hypothesis relies heavily on the premise that a neutral species, in this case $N_2$, can escape from the high-field region of the tip without undergoing ionization. Since direct measurement of the proposed neutral species is not possible with current APT instrumentation, it cannot be proven nor disproven from the existing data.

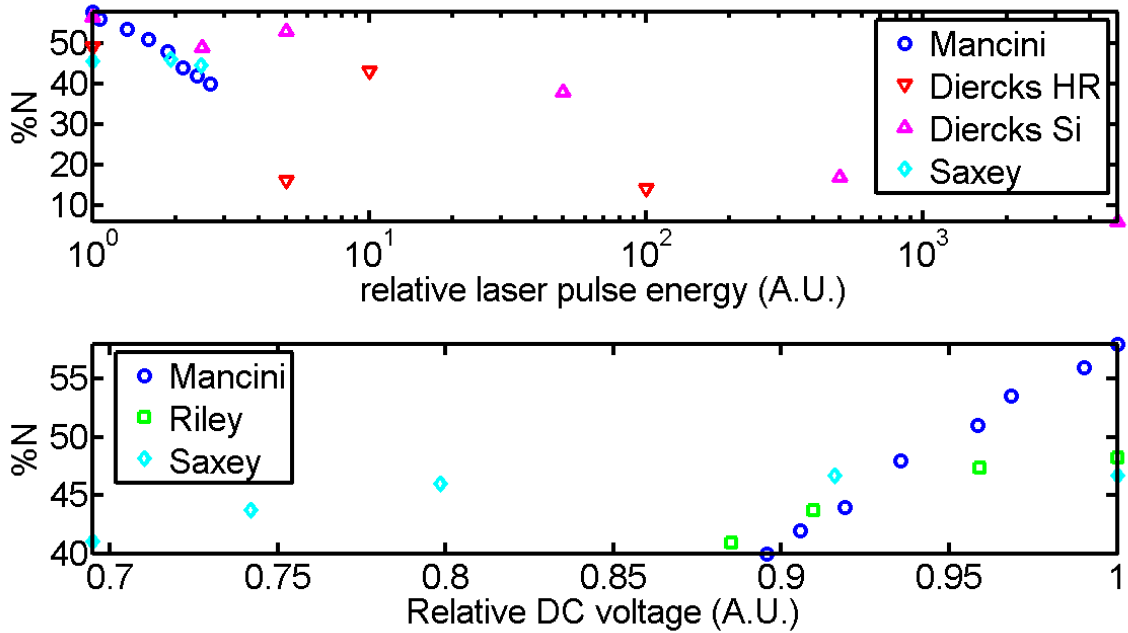

*Figure 2: Evolution of the N concentration in the analysis of GaN as reported by Mancini et al. [63], Riley et al. [57], Diercks et al. [55] and new data processed by Saxey based on data from ref. [64]. (top) as a function of the laser pulse energy relative to the lowest energy reported in the article and (bottom) as a function of the standing voltage relative to the highest voltage reported in the article. Mancini, Riley and Dierks Si are obtained on straight flight-path atom probes. Saxey and Dierks HR are obtained on reflectron-fitted instruments. The trends are clearly similar for these different*

*studies.*

### B. THERMAL DESORPTION OF $N_2$ FROM GaN

The thermal stability of GaN has been extensively studied. Ambacher et al. reported that the emission of $N_2$ from thermal decomposition of GaN grown by metal-organic chemical vapor deposition follows an Arrhenius rate law at temperatures above 1000 K [65], with no noticeable emission of $N_2$ at lower temperature. Similar trends are observed for InN and AlN at temperatures above 900 K and 1300 K, respectively [65]. The activation energy for thermal desorption of $N_2$ from GaN is 3.93 eV, while the pre-exponential term is $1.2 \times 10^{35}$ $m^{-2}$ $s^{-1}$, in good agreement with other values found in the literature [66–68].

To model the influence of thermal desorption in the case of APT, we can assume that the heated volume at the apex of the specimen can be approximated as a cylinder 10 μm in length capped with a 50 nm radius hemisphere. The thermal pulse can be modeled as a 1 ns square pulse with a repetition rate of 500 kHz. It is reasonable to expect that the surface of GaN will not be heated above 600 K at the peak of the thermal pulse. Based on an integration of the Arrhenius equation for the desorption rate, the thermal desorption flux can be deduced: $1 \times 10^{-13}$ $s^{-1}$. In comparison, the ion collection rate in a typical APT analysis, ranges between 1500 $s^{-1}$ and 5000 $s^{-1}$. This would be insignificant over the course of any atom probe experiment. If the temperature was to rise to 1000 K, the rate increases to 2 $s^{-1}$. Based on these calculations, a significant loss of $N_2$ through thermal decomposition of GaN seems unlikely in the context of APT experiments.

Several aspects should however be considered. In the case where very high temperatures, above 1000K, were reached subsequent to the absorption of the laser pulse as was reported for other semiconductors [46], then the rate could become competitive with the field evaporation rate. Other processes, particularly surface diffusion of the Ga and N on the GaN surface would however probably play a significant role and the data quality would be affected. More importantly, one could assume that first a molecule of $N_2$ forms and then desorbs from the surface. The activation energy used above would then not necessarily be relevant, as indeed, the desorption barrier will be affected by the presence of the strong electric field necessary for APT.

### C. EFFECT OF THE ELECTRIC FIELD ON $N_2$

The effect of the electric field on gas atoms or molecules near a field emitter has been studied in the context of field-ion microscopy (FIM)[69]. FIM is a predecessor of APT, utilizing similar specimen geometries and electric field intensities. In FIM, mono-atomic gases (e.g. He, Ne, Ar) are typically used to produce a highly-magnified image of the surface of an emitter via field ionization near the specimen surface [69]. Due to polarization forces, molecules will migrate *towards* the highest field region at the apex of the specimen in successive hops, bouncing off the surface, and progressively losing kinetic energy in a process referred to as thermal accommodation. Eventually, the atoms cross a region where the electric field is sufficient to cause their ionization and projection towards a detector screen. The ionization of imaging gas atoms, as detailed in ref. [70], is affected by the presence of a layer of adsorbed imaging gas at the specimen surface which reduces the activation energy for field ionization and facilitates the accommodation process [71,72]. The binding of such adsorbed atom or molecule is due to a chemical effect and a polarization effect. The latter is referred to as a dipole-dipole interaction, of higher magnitude than the force originating from the volume polarization, and is linked to the local enhancement of the electric field at emitting sites on the surface.

If $N_2$ molecules have desorbed from the tip during an APT experiment, these gaseous molecules should behave comparably to the imaging gas in FIM. As detailed in Table 1Table 1, $N_2$ has a volume polarizability of the same order of magnitude as Ar, and higher than that of Ne. The work by Suchorski et al. [73] also showed that $N_2$ behaves similarly to mono-atomic gases with respect to imaging in FIM, with a critical electric field of 17 V nm$^{-1}$ to induce $N_2$ ionization. Tomanek et al. [74], who investigated the specific case of $N_2$ on a metallic Fe (111) surface, derive a value of the electric field that will directly lead to desorption and ionization of $N_2$ from the surface of 15 V nm$^{-1}$ in good agreement with the measurements of Suchorski et al. [73] (17 V nm$^{-1}$), and they reported total field adsorption binding energies ranging from 1.5 to 5 eV. Collectively these studies suggest that gaseous $N_2$ should act like any other FIM imaging gas.

| Atom or Molecule | volume polarizability ($\alpha/4\pi\varepsilon_0$) ($10^{-3}$ nm$^3$) |
|---|---|
| O | 0.802 |
| $O_2$ | 1.56 |
| N | 1.1 |
| $N_2$ | 1.71 |
| He | 0.204 |
| Ne | 0.392 |
| Ar | 1.63 |

*Table 1: Volume polarizabilities of selected atoms and molecules. Data from the Computational Chemistry Comparison and Benchmark DataBase (http://cccbdb.nist.gov/).*

GaN is generally stable at the base temperature considered here, and no noticeable increase in the overall pressure in the ultra-high vacuum chamber of the microscope is observed over the course of the experiment. So if $N_2$ was to be thermally desorbed during the course of the experiment, it would do so in low quantities. In addition, conversely to the case of FIM where atoms from the imaging gas need to reach the high field region near the apex of the specimen, in this case the $N_2$ would be in the vicinity of the surface already. Ultimately, assuming that neutral $N_2$ could still be thermally desorbing from the surface, and escaping, as a neutral particle with enough energy to travel away from the surface and remain unionized, the applied field eventually drives gaseous species (adsorbed or desorbed) towards the highest electric field region of the tip rather than away from it.

At low pressure, it is likely that the N2 molecules will only form the adsorbed layer at the surface. Assuming that the case of $N_2$ above a GaN surface is equivalent to He on W, and with a field enhancement factor of 2 at emitting sites, we can estimate that the polarization-energy contribution to the total binding energy at a field of 20 V nm$^{-1}$ is in the range of almost 1 eV, which is at least an order of magnitude larger than the energy of the thermal agitation of the surface atoms even at 1000 K. Ultimately, it is this rather unlikely that $N_2$ should be lost to the surrounding vacuum through desorption from thermal agitation, and altogether lost from the analysis, even if its contribution was part of the background.

### D. IONISATION OF N AND $N_2$ ABOVE THE GaN SURFACE

To investigate the probability of ionisation of N and $N_2$ above a GaN surface, we make use of the model developed by Kingham for visualizing the post-ionization theory [29,75]. The model allows for prediction of the ratio of charge-states of the ions as a function of the applied electric field. There are no assumptions made within Kingham's framework that preclude adapting it to the case of Ga, N and $N_2$ ionisation above a GaN surface. This model has been subsequently supported by experimental observations on numerous occasions [75–77], and although it was shown to fail to predict the precise charge-state in a case where molecular ions were formed, its results were qualitatively satisfactory [53]. The implementation of the equations are the ones developed by Yao et al. [78] The work function of GaN, and the successive ionization energies of Ga, N and $N_2$ that are listed in Table 2 have all been utilized as inputs.

|                | Ga    | N     | $N_2$ | $O_2$ |
|----------------|-------|-------|-------|-------|
| **1st Ionization** | 6.00  | 14.55 | 15.58 | 12.07 |
| **2nd Ionization** | 20.54 | 29.64 | 43.8  | 16.1  |
| **3rd Ionization** | 30.75 | 47.51 |       |       |

*Table 2: Values of the ionization energies in eV for Ga, N and $N_2$ were taken from ref.* [79–81].

The resulting curves are displayed in Figure 3~~Figure 3~~. Several features are worth highlighting: (i) the value of the ionization field of $N_2$ is in good agreement with the experimental value reported by Suchorski et al. [73]; (ii) the distribution for Ga is close to the one obtained by Kingham [29]; (iii) double ionization of $N_2$, as expected from its very high second ionization energy, can only occur at extremely intense electric fields well above 60 V nm$^{-1}$, so at fields where we should expect to exclusively observe $Ga^{3+}$ which is opposite to experimental measurements reported by i.e. Mancini et al.. The results presented herein indicate that detecting $N_2^{2+}$ in the analysis of GaN is extremely unlikely, conversely to what has sometimes been reported [63]. This means that a discrepancy in composition measurements cannot be attributed to $N_2^{2+}$ potentially being observed at 14 Da in the APT mass-to-charge-state-ratio spectrum and mistakenly identified as $N^+$. Indeed, based on energetic considerations, the formation of $N_2^{2+}$ is even less likely than $O_2^{2+}$, the absence of which had been proven by isotopic enrichment experiments in both Fe oxide [82] and Si oxide [83].

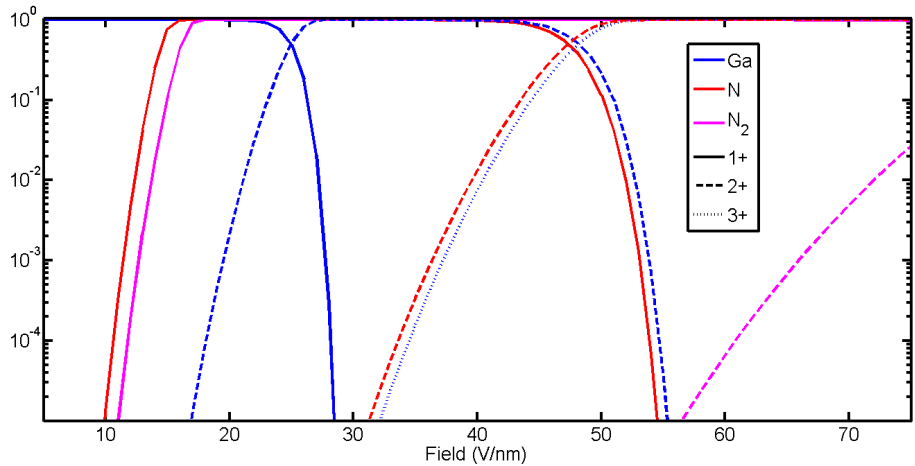

*Figure 3: Kingham curves displaying, as a function of the electric field, the relative abundance of each charge state for Ga, N, and $N_2$ above a GaN surface.*

## III. MOLECULAR IONS IN HIGH FIELDS

### A. DISSOCIATION OF MOLECULAR IONS

Field evaporation of molecular ions has long been documented [84], and was attributed to a lower critical electric field to induce field evaporation associated with molecules rather than individual atoms. Such particles can be formed either by one or more mobile atoms adsorbed on the surface [85] bonding with a surface atom or by atoms that are neighbors within the structure of the specimen. During field evaporation, the desorbing molecule becomes charged and can encounter field-dissociation. It has previously been confirmed that this process can occur, and has in fact been studied in the case of metal-halides and hydrides [86,87].

Recently, evidence for the dissociation of molecular ions in the APT analysis of GaN was provided by Saxey [50], who introduced the concept of correlation histograms to investigate the behavior of pairs of ions detected near-simultaneously. Similar histograms have been employed in mass spectrometry to study the dissociation behavior of small molecules in intense fields [88]. Figure 4Figure 4(a) shows one such correlation histogram. The respective mass-to-charge-state ratios are plotted for the case when the detection of two ions was correlated to the same laser pulse. These histograms reveal potentially non-random combinations of mass-to-charge-state ratios that can be directly related to physical processes occurring at the time of, or shortly after, field evaporation. Of particular interest here are tracks corresponding to the dissociation of molecular ions that are highlighted in the figure by thin grey lines. Based on the locations of the origin and the terminus of each track, it is possible to deduce the dissociation reaction. For example, tracks shown in Figure 4Figure 4(a) correspond to the dissociation process $GaN^{2+} \rightarrow Ga^+ + N^+$. The difference in the respective mass-to-charge-state ratios, as measured from the origin of the track, provides information on the intensity of the electric potential at the location of dissociation. This was termed the 'voltage drop' [50], although more accurate would be 'potential drop'. The potential drop can be used to calculate where and when the dissociation took place along the ion's trajectory from the specimen to the detector. Additionally, it is evident from the faint track shown in Figure 4Figure 4(b) that a dissociation reaction took place that led to the formation of a neutral N₂: $GaN_3^{2+} \rightarrow Ga^+ + N^+ + N_2$. These molecular dissociations are currently the only manner in which neutral species can be detected directly by APT measurements.

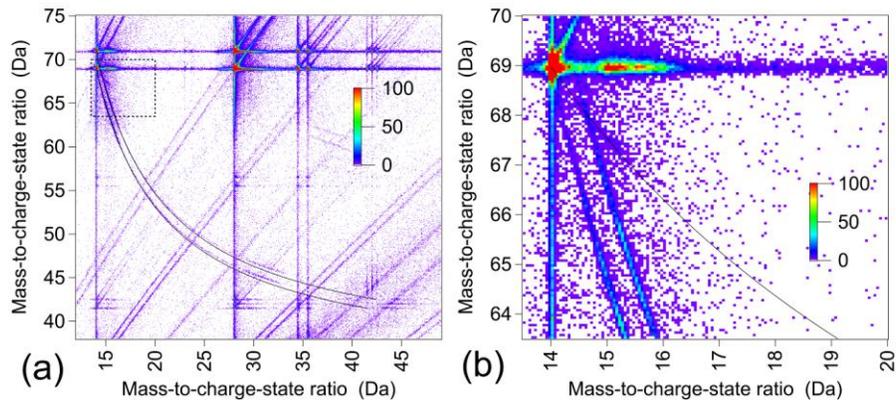

Figure 4: (a) Correlation histogram for GaN data presented in ref. [50]. (b) close-up on one of the dissociation tracks that corresponds to a dissociation producing a neutral $N_2$ highlighted.

| Parent | | Fragments | | | | | Delta E (eV) |
|---|---|---|---|---|---|---|---|
| **GaN²⁺** | -> | Ga⁺ | + | N⁺ | | | -1.63 |
| **GaN³⁺** | | Ga²⁺ | + | N⁺ | | | -9.94 |
| **GaN₃²⁺** | -> | Ga⁺ | + | N₃⁺ | | | -7.86 |
| **GaN₃²⁺** | -> | Ga⁺ | + | N⁺ | + | N₂ | -3.98 |

Table 3: Dissociation energies for a selection of molecular ions observed experimentally.

For those dissociation paths, as well as those reported in ref. [50], we used DFT simulations to derive the dissociation energies of $Ga_nN_m$ cations, as reported in Table 3Table 3. The calculated values are in good agreement with those reported by Kandalam and co-workers [89]. Interestingly, these molecular ions are not stable and their dissociation is energetically favorable. These simulations do not account for the effect of the electric field, which can affect the relative stability of some molecular ions as discussed by several authors for the case of formation of $H_3^+$ ions [90,91], which could explain how such molecular ions are found field evaporated in this form. The case of the dissociation reaction for $GaN_3^{2+}$ is particularly interesting, as there is a dissociation path that leads to the formation of a neutral $N_2$ molecule, which is seen experimentally as shown in Figure 4Figure 4(b). This also reveals that the most common dissociation path leads to the formation of $N_3^+$. Finally, such dissociations are known to also take place in the APT analysis of other material systems, for example, Santhanagopalan et al. [92] reported the formation of a neutral $O_2$ molecule originating from the dissociation of a $PO_4^+$ molecular ion.

## B. FURTHER EVIDENCE OF NEUTRALS FROM DISSOCIATIONS

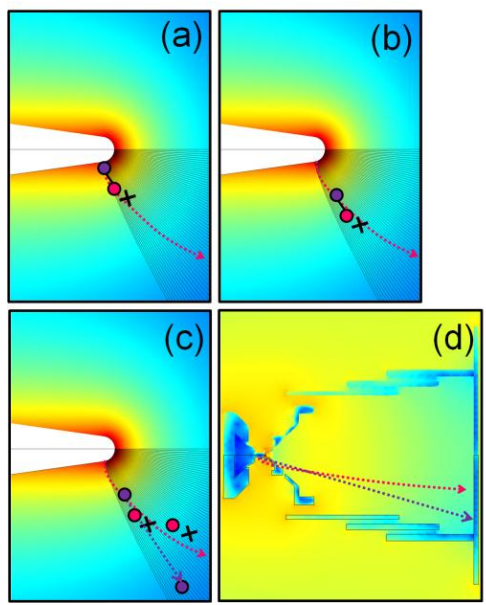

*Figure 5: Schematic illustration of the difference in expected trajectory for a neutral and charged fragment: (a) the molecular ion is emitted from the surface; (b) the ion reaches the point of dissociation; (c) the charged fragment (pink) follows a curved trajectory while the neutral fragment (purple) keeps flying straight; (d) perspective on a full-scale microscope.*

We now aim to provide further experimental evidence that neutrals can be generated by the dissociation of molecular ions, since proofs of their existence have been rather indirect so far, as previously pointed out by Tsong in section 2.4 of ref. [93]. In their study of GaSb, Müller et al. [32] utilized the analysis approach developed by De Geuser et al. [94] to plot a histogram of the distance separating the relative positions of two ion hits on the position-sensitive detector specifically for the case where these ions were generated either by the same pulse or by two successive pulses. These histograms exhibited an unexpected hump at small distances, which was attributed to ions originating from molecular ion dissociations. This is a likely hypothesis considering the high proportion of multiple events (i.e. when multiple detected ions are correlated to the same laser pulse) in the analysis under these high electric field conditions. The analysis indicates that charged fragments, formed by a single dissociative event, arrive separated by close distances on the detector.

This can be explained by the fact that ions follow well-defined trajectories that are dictated by the static field distribution [95], notwithstanding their mass or charge. It is expected that two charged fragments approximately follow the same path since they originate from the same location at the specimen surface. In contrast, the respective trajectories followed by a charged and neutral fragment pair resulting from the same dissociation event should diverge significantly. Momentum is initially acquired by the neutral fragment while it belongs to the parent ion. At the point of dissociation, it can be assumed that it follows a straight trajectory that is the local tangent to the instantaneous trajectory of the parent ion. In contrast, the trajectory of the charged fragment is defined by the static field distribution. These scenarios are depicted schematically in Figure 5Figure 5.

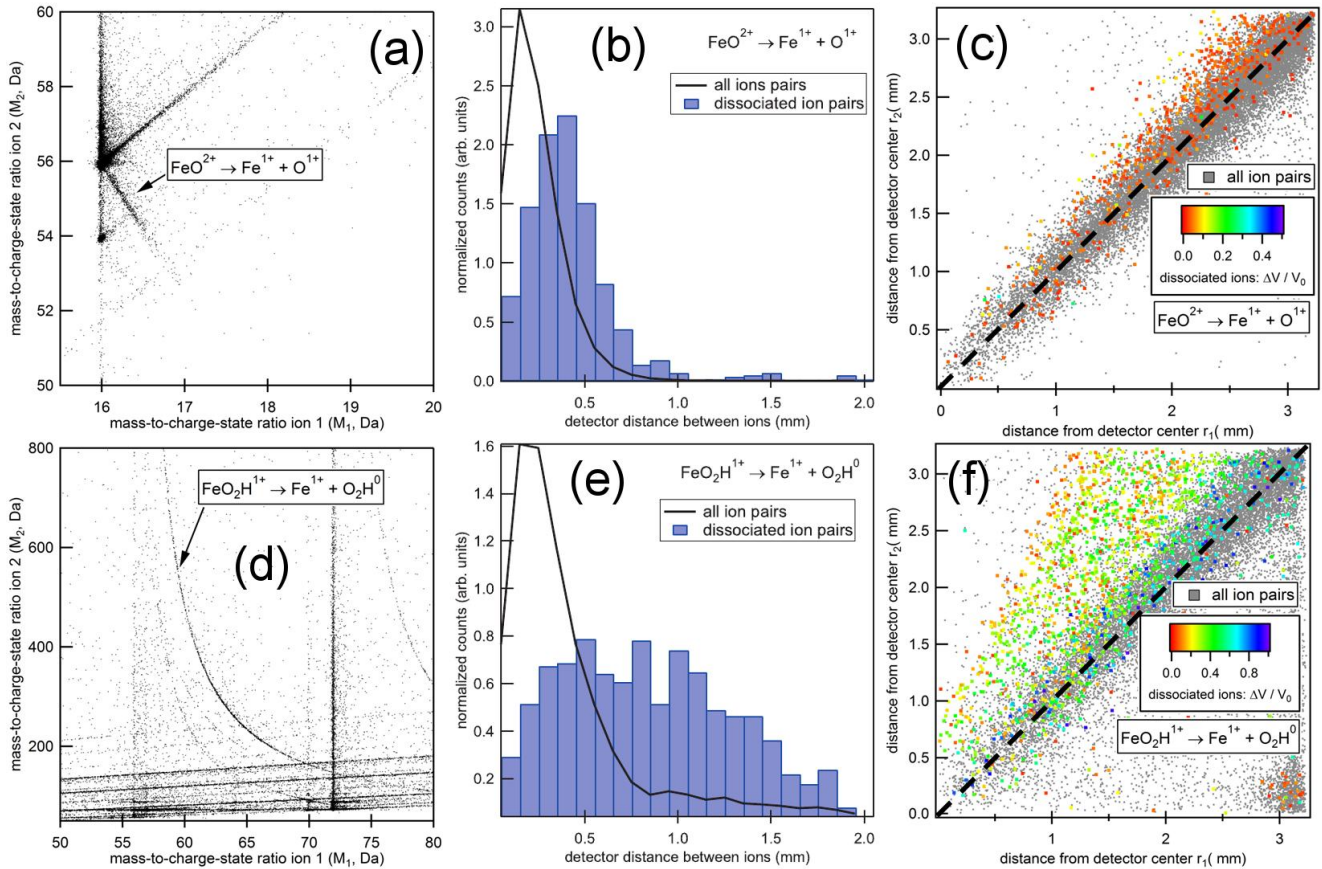

*Figure 6: (a) section of a correlation histogram for a single $Fe_3O_4$ analysis in the (111) orientation showing well-defined tracks corresponding to the dissociation of $FeO^{2+}$. (b) Histogram of distance between hits for all pairs coming on multiples hits and only those from the dissociation track. (c) Graph of the distance to the centre of the detector for all hits (in grey) and for hits corresponding to dissociation tracks color-coded based on the energy loss. The dashed line. a $r_1=r_2$ is provided as a guide to the eye. (d) Section of the correlation histogram from the same analysis showing a dissociation leading to a neutral $O_2H$. (e–f) Similar graph as in (b) and (c) respectively for hits corresponding to a dissociation track involving a neutral particle, color-coded based on the energy loss.*

Figure 6Figure 6 (a) shows a section of a correlation histogram centered on a track corresponding to a dissociation reaction with only charged products, i.e. $FeO^{2+} \rightarrow Fe^+ + O^+$. Figure 6Figure 6 (b) is a histogram of the distance between the detector positions measured for each ion in a pair of hits registered as a multiple event. All the multiple events are plotted as a black line, pairs specifically from the dissociation track are showed as a bar plot in blue. The thin black line shows a strong peak at short

distance indicating that ions on a multiple hit tend to arrive close together on the detector, indicating that they followed very similar trajectories and originated from close loci at the specimen surface. This is similar to observations reported earlier [32,94]. The distribution of distance for ions on the dissociation track is rather similar, if slightly shifted towards larger distances. Figure 6 (c) plots the respective distance to the center of the detector measured for each ion in a pair of hits registered as a multiple event. All the multiple events are plotted in grey, pairs specifically from the dissociation track are highlighted in color. The distributions approximate a straight line ($r_1=r_2$). This was expected considering the distribution in Figure 6 (b) with ions from multiples events hitting the detector at short distance from each other. In this case, the distribution for ions generated by the dissociation is comparable, implying that the trajectories of a pair of ions that are the product of dissociation are similar to that of a pair of ions that have been individually field evaporated from the tip. The color-code in the figure relates to the relative potential drop. Dissociations occurring early in the trajectory from the specimen to the detector are shown in red, whereas dissociations taking place later during the flight are shown in blue. Most of the dissociations in this case took place early during the flight.

Let us now consider a dissociation reaction that results in the formation of a neutral fragment, $FeO_2H^+ \rightarrow Fe^+ + O_2H$ indicated in Figure 6 (d). The histograms of distance between ions from a multiple event for all ions pairs and pairs originating from dissociations are shown in Figure 6 (e). The slight difference in the distribution labeled *all ion pairs* in Figure 6 (b) and (e) is due to the track fitting routine that has filtered down hits to only consider those with mass-to-charge-state ratio within 5 Da of the track's minimum or maximum mass-to-charge ratio so as to ensure a meaningful comparison. Here, the distribution changes significantly, as shown in Figure 6 (e). This distribution exhibits a much wider spread, with a shift of one of the dissociation products (the neutral fragment) towards larger distances from the projection centre. The difference between the two distributions in this case is striking, indicating that the distance between a charged and a neutral fragment is on average much larger than when the two fragments are charged. Figure 6Figure 6 (f) is a similar plot as in is Figure 6Figure 6 (c), with, again, the color-code that relates to the relative potential drop. There is a wider spread in potential drop than exhibited in Figure 6Figure 6 (c). In this case, the earlier the dissociation occurs (red) the farther are the two impacts on the detector, while late dissociations (blue) tend to lie on the $r_1=r_2$ line.

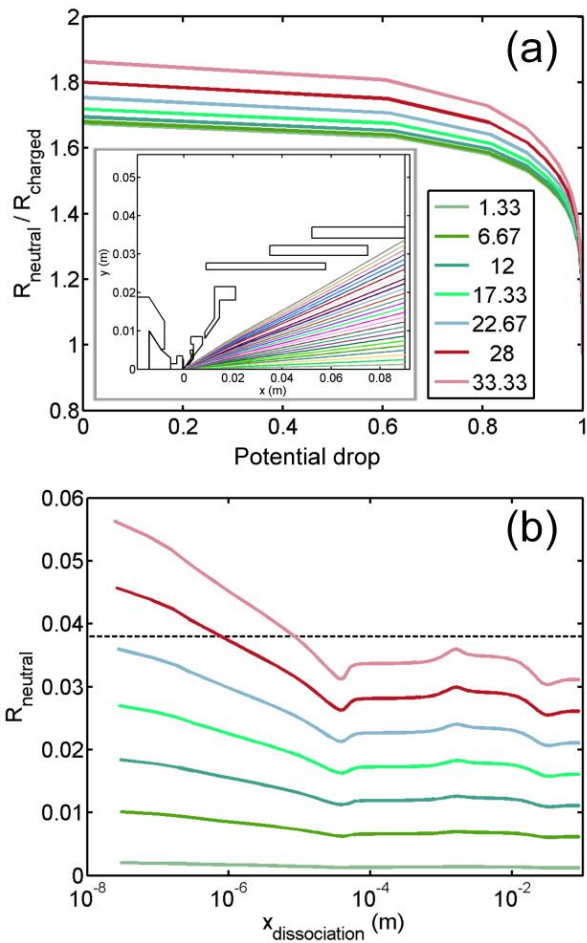

Figure 7: (a) Relative impact position for a neutral fragment compared to a charged fragment as a function of the potential

*drop, with curves of different colours corresponding to different launch angles (as reported in the legend). The full set of trajectories from this particular simulation is shown in the inset. (b) Absolute impact position for a neutral fragment as a function of the dissociation position. The dashed black line highlights the position of the physical detector boundary.*

To assess the validity of these results, we have made use of complementary simulations of ion trajectories within a full-size atom probe microscope. These simulations were undertaken by Loi et al., and are presented in detail in ref. [26]. A set of trajectories was computed for the case of a sample mounted on a microtip array, typical of an APT specimen prepared by focused ion beam [96] The simulated specimen incorporated a shank angle of 10° and radius of curvature of 70 nm, which is close to the experimental case for the APT analyses reported in Figure 6. Analyses of the resulting trajectories are presented in Figure 7(a). The detector hit position of a neutral fragment, assuming that it flew straight from the point of dissociation relative to its charged counterpart, is reported as a function of the potential drop. In Figure 7 (b) we report the actual detector hit position of the neutral product as a function of the position of the dissociation along the x-axis, the specimen-to-detector direction, within the APT analysis chamber.

These graphs indicate that as the molecular ion dissociates earlier (closer to the specimen) during its flight, i.e., at smaller *x* position and lower potential drop, the landing location of the neutral fragment is further away from the centre of the detector. This is because small *x* positions correspond to a stage in the ion flight where its trajectory is still highly curved. In addition, there is a minimum *x* position threshold for the dissociation below which a neutral fragment may become completely undetected, as the resulting lateral trajectories will fall outside the outer edge of the detector (detector radius is typically 38 mm in commercial instruments). There are assumptions in this treatment, in particular, regarding possible rotational degrees of freedom for the molecular ion, which may further affect the trajectories of the neutral fragment [97]. In addition, the meshing of the simulations is probably too coarse, especially for ion positions close to the specimen, to allow for an accurate quantitative analysis. However, these results are qualitatively in close agreement with the experimental observations and help to confirm that neutral fragments are effectively generated by the dissociation of molecular ions. Indeed, it is a concern that compositional analysis will be affected by such processes dependent on the initial position of the molecular ion at the specimen surface.

## IV. DISCUSSION

The new results presented in this study contrast significantly with the mechanisms proposed in the recent literature. It appears unlikely that molecular $N_2$ can be thermally desorbed from the specimen surface during an APT experiment without undergoing subsequent ionization. In cases where the surface was to reach extreme temperatures, in excess of 800 K or 1000 K, then thermal desorption could be possible. Without any other source of initial velocity to promote the departure of $N_2$, nor any further acceleration to facilitate escape from the high-field region of the tip, thermal desorption is not energetically favorable and its potential impact on the experimental results minimal. The key physical parameters are the nitrogen binding energies, the holding-field value used in APT, and the successive ionization energies of the different species. Furthermore, there is some convincing evidence that , under the operating conditions of laser-pulsed atom-probe tomography, the surfaces of semiconductors mimic metal surfaces, which reinforces the validity of our approach [49,98]. In the range of electric field intensities implied by the Ga charge-state ratios reported by Mancini and co-workers [63] of 20 V nm$^{-1}$ to 27 V nm$^{-1}$, N and $N_2$ are both expected to be entirely singly ionized (i.e., $N^+$ and $N_2^+$ only, with no other charge states). In ref. [99], de Castilho et al. provide an analytical formula for a one-dimensional distribution of the electric field as a function of the distance along the specimen axis . For a specimen with a radius of curvature of 50 nm and an electric field of 20 V nm$^{-1}$ at the surface, the electric field remains above 17 V nm$^{-1}$, so sufficient to ionize $N_2$ and N, for a distance of nearly 5 nm. Without some form of acceleration or initial velocity, it also seems unlikely that a potentially desorbed neutral molecule could drift such a distance without ionization. Furthermore, an escaping polarized molecule would eventually be drawn back towards the specimen apex, and should hence remain in the vicinity of the specimen. The assembled evidence casts significant doubt on the viability of a loss mechanism that involves the direct desorption of neutral molecules in the case of GaN.

There is no reason why the application of APT to oxides should be any different. It was recently reported that discrepancies in the APT measured composition of different oxides could be explained by either the formation of negatively charged dioxygen ($O_2^-$) migrating down the shank of the specimen or desorption of neutral $O_2$ directly from the surface [100]. The formation energy of a $O_2^-$ anion is in the range of 15 eV [101] in the gas phase, which is high. Thus, it is expected to be extremely difficult to form $O_2^-$ at a surface bearing a very high positive charge, corresponding to an electric field in the range of $10^{10}$ V m$^{-1}$ or more. In addition, the migration of adatoms away from the apex is not supported by experimental evidence. Indeed, observations of field-driven adatom migration [85,102] have shown that adsorbed species migrate towards maxima of the highest electric field. The polarizability of gaseous $O_2$ is in the same range as $N_2$ and Ar, as presented in Table 1. Thus, in the presence of a high electric field, the behavior of $O_2$ and $O_2^-$ is expected to be similar to the behaviors of $N_2$ and Ar, which are attracted towards higher electric fields in the vicinity of the apex. Finally, the first ionization energy of $O_2$, as reported in Table 2, is lower than that of $N_2$. Thus, the likelihood that $O_2$ would be ionised in an electric field in the range of 20 V nm$^{-1}$ cannot be ignored.

How then can compositional errors in the APT analysis of these materials be explained? The energy barrier for field evaporation is indeed species-specific [103,104], and likely to also be specific to the local atomic neighborhood for a given element [105,106]. Hence, so too is the probability that an atom can field evaporate. For example, under certain conditions this can lead to the field evaporation of specific species in the standing field, independent of any applied high voltage- or laser-pulsing. Such uncorrelated evaporation precludes time-of-flight identification, and generally leads to these ions being relegated to contributions to the background in the mass spectrum. It should be noted, however, that the background level is lowest in both GaN and various oxide datasets when the N or O deficiency is at its highest (i.e., low field and high laser fluence). This point has been emphasized in previous studies as supporting evidence for the direct loss of neutrals being a major contributor to measured stoichiometry deficiency [55,58]. If these observations are not the result of direct neutral loss, and are inconsistent with simple selective ionization loss into the background, an alternative explanation is needed.

Molecular dissociation is increasingly likely to be a possible explanation. It is clear from experiments that molecular ions are observed in field evaporation. Those may however not be energetically stable as suggested by the results reported in Table 3Table 3. As they fly away from the specimen, molecular ions can dissociate into charged or neutral fragments. We have demonstrated that neutrals are effectively detected in APT and are typically discarded as contributions to the background, as discussed in ref. [50]. In cases where the field is high (low laser fluence), the dissociation is more likely to occur and to do so in a region, close to the specimen surface, where the neutral fragment can undergo field ionization. In such a case, the ion is accelerated and will therefore be detected as part of a multiple event. For compound semiconductor, the proportion of multiple events is known to increase as the laser fluence decreases [32], thereby supporting this hypothesis. At lower field (high laser fluence), dissociative events could be occurring far enough from the specimen surface such that the intensity of the field is too low to induce subsequent ionization of the neutral fragment. How far is enough depends on the natures of the neutral and the specimen, as well as the specimen and microscope geometries, but if the dissociation does not occur in the early stages of the fly, it is very unlikely that a neutral fragment will be ionized and it will thus be excluded from the analysis.

As previously discussed, quantitative analyses of the intensity of the tracks, observed in the correlation histograms presented by Santhanagopalan et al. [92] in the analysis of LiFePO$_4$, did not result in a significant difference in the amount of O. In that study, molecular dissociations accounted for only 1.6 % of the ions, increasing the atomic oxygen concentration by 0.3 %. Detection of neutral species from molecular dissociation is also complicated by two significant factors. First, the majority of atom probe microscopes currently in operation are equipped with a reflectron lens., which is used to improve the mass resolution of the instrument. A reflectron is a concave electrostatic mirror that bends the ionic trajectories, and, in commercial instruments, the ions are deflected to an angle in the range of 120° to 180°. However, any neutral formed prior to entering the reflectron will not be affected by the electrostatic field and will never reach the detector.

Second, in the case of atom probes without reflectrons, a second— possibly more significant—issue arises. Neutral fragments formed by dissociation were only accelerated during the time they were part of the original parent ion. Hence a neutral fragment gains (in its total flight) less kinetic energy than it would have gained if emitted as a charged entity at the emitter. The current generation of single ion detectors utilizes microchannel plates (MCPs) to amplify the signal from a single ion to thousands of electrons. Typically, the detection efficiency of the MCP, i.e., the fraction of ions striking the detector that are actually registered as hits, is proportional to the open area of the detector (≈ 60 %). However, this efficiency also decreases at small ion energies. Gao et al. showed a consistent relationship between MCP efficiency and ion energy that is largely independent of ion mass [107] and that this decreases sharply below ion energies of ≈ 2 keV. Ordinarily, this effect is not significant, because APT experiments are typically performed at voltages well in excess of 2 kV. However, there are implications for APT measurements if neutrals formed by dissociation acquire, prior to formation, only some fraction of the kinetic energy corresponding to the full extraction potential $V_0$. To further consider this possibility, we revisit the dissociation track in Figure 6(c) that results in a neutral $O_2H$ fragment. So as to facilitate quantification, and following Santhanagopalan et al. [92], the data are first replotted in Figure 8 (a) as the minimum distance of each ion pair from the dissociation track as a function of the potential drop. A histogram quantifying the dissociation curve as a function of potential drop is plotted in Figure 8(b) (blue bars). Next, applying the relationship from Gao et al. to the case of $O_2$ neutrals formed from dissociation of molecular complexes, the expected MCP efficiency is calculated as a function of the potential drop for an assumed operating voltage (6.6 kV for these particular data) and is shown as a dashed line, as shown in Figure 8(b). It is apparent that when the relative potential drop is less than ≈ 0.2, the $O_2$ neutral is significantly less likely to be detected. Applying the calculated detection efficiency to the quantification of the dissociation track, it is possible to partially correct for the decreased detection efficiency, the corresponding histogram of which is plotted in Figure 8(b) (red bars). In this case, correction increases the apparent number of $O_2$ neutrals in the histogram by approximately a factor of 2, highlighting the significant loss of $O_2$ neutrals due to the energy-dependent MCP detection efficiency. It should also be noted that the MCP detection efficiency cutoff for low-energy molecules generally occurs before a time-of-flight window cutoff for the detection (based on the time between successive pulses) in these data. Hence, collecting data for example at a slower pulse repetition rate (larger time-of-flight window) would unfortunately not increase the detection of these types of dissociations.

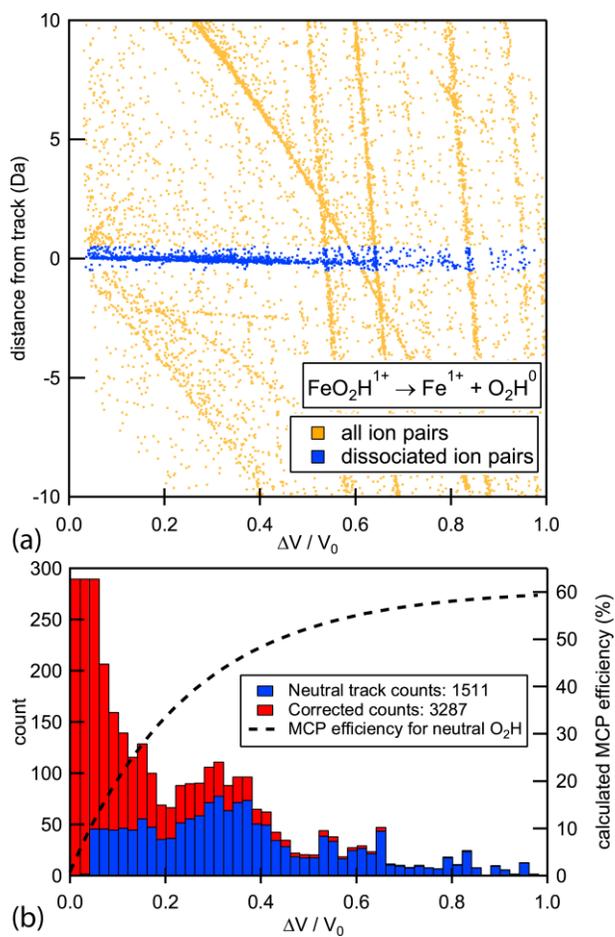

*Figure 8: (a) Dissociation track of Fe-O molecular complex leading to a neutral daughter replotted for quantification. (b) Quantification of the same track (blue) as a function of $\Delta V/V_0$ and corresponding calculated MCP detection efficiency for neutral $O_2$. Red histogram applies the expected detection efficiency to approximately correct for MCP efficiency losses.*

This could be particularly problematic for evaporated molecular species that are highly unstable. This might suggest that the detected neutral fragments originate from relatively *stable*, but still metastable, molecular ions that do not dissociate until slightly after initial ionization. This hypothesis is supported by additional *ab-initio* calculations corresponding to the data presented in Figure 6Figure 6. As reported in Table 4Table 4, metastable molecular ions, such as $FeO^{2+}$ or $FeO_2H^+$, are observed to dissociate in the APT experiment. Their respective energy of dissociation, derived from DFT calculations, highlight that they are rather unstable, and that FeO2H+ is relatively more stable, which could explain why the dissociation occurs at a later stage during the flight. The relative stability of some molecular ions, their probability of dissociation, and the possibility that some neutral fragments go undetected, can explain oddities in the trends of measured charge states of Fe ions in magnetite that deviate from post-ionization theory [53].

| **Parent** | | **Fragments** | | | **Delta E (eV)** |
|---|---|---|---|---|---|
| **$FeO^{2+}$** | -> | $Fe^+$ | + | $O^+$ | -3.71 |
| **$FeO_2H^+$** | -> | $Fe^+$ | + | $HO_2$ | 4.59-0.79 |

Table 4: energetics of the dissociations observed in Figure 6Figure 6.

Overall, the current analyses identify two new explanations for the stoichiometry deficiencies. Firstly, a neutral fragment that forms early in flight will not follow a conventional ion trajectory, possibly resulting in an impact outside the detector field of view, while a charged fragment that forms early in flight will arrive as normal at the detector. Secondly, even if the neutral fragment impacts the detector, it is likely that it may not have sufficient energy to trigger the necessary electron cascade on the MCP to be considered an ion impact, thereby precluding any contribution to the background and even more so its elemental identification.

# V. CONCLUSIONS

Collectively, the theoretical and experimental results reviewed in this study directly address critical flaws and outstanding questions in the current literature on the behavior of molecular ions and neutral molecules in intense electric fields, particularly regarding the hypothesis of direct thermal desorption of neutral molecules under the high field conditions encountered during an atom probe experiment.

In summary:

- an array of evidence has been provided, based on the field ionization and post-ionization theories, demonstrating that it is very unlikely that dinitrogen can remain neutral in the vicinity of GaN surfaces under the electric fields typically experienced in the APT analysis (which would be unlike the behavior of other gas molecules used in field ion microscopy);

- direct experimental evidence of molecular dissociation leading to the formation and detection of neutral fragments away from the impact of charged fragments has been presented. This evidence is further supported by boundary-element method simulations;

- DFT calculations have been implemented to perform energetic calculations in order to support experimental observations of molecular ion dissociation;

- we proposed that discrepancies observed in APT analyses of nitrides and oxide originate from $N_2$-carrying or $O_2$-carrying molecular ions dissociating during the flight, where the electric field is insufficient to induce ionization. The neutral fragment will then have acquired sufficient kinetic energy to escape the vicinity of the charged specimen but its trajectory may never intersect the detector, or it may not have enough energy to be detected.

  s

## VI. METHODS

Specimens for APT analysis were prepared by focused ion-beam, using a conventional lift-out procedure[96], while some of the specimens depicted in Figure 1 were nanowires of GaN requiring no focused ion beam processing. Data presented in Figure 2 and 3 were acquired on different commercial instruments from Cameca Instrument. Please note that commercial equipment, instruments, or materials are identified only in order to adequately specify certain procedures. In no case does such identification imply recommendation or endorsement by the National Institute of Standards and Technology, nor does it imply that the products identified are necessarily the best available for the purpose. The instruments used were: LEAP 3000 and 4000 HR or Si and LAWATAP. Details of the experimental conditions are presented in the cited publications. Data presented in Figure 4 were acquired on a Cameca LEAP 3000X HR equipped with a green laser; see details in ref. [50]. Data presented in Figures 6 and 8 were collected with a LEAP 4000X Si at 40 K, with a 355 nm wavelength laser operating at 1.2 and 40 pJ/pulse and a detection rate of 0.0045 detected ions/pulse. Further details can be found in[53].

For the DFT calculations, we optimized the geometries of all small molecules using density functional theory and B3-LYP hybrid functionals with the standard 6-311G basis set. The final energy of each molecule was then computed with a single-point calculation using the coupled cluster method with quadruple-zeta basis sets augmented with diffuse functions. All calculations were performed using the Gaussian09 software package [108].

## VII. ACKNOWLEDGEMENTS

BG acknowledges Shyeh Tjing (Cleo) Loi (University of Sydney) who performed the BEM

simulations, with support from Drs Brian Geiser & Drs David Larson (Cameca). Dr Frederic De Geuser (CNRS, UJF, France) is acknowledged for fruitful discussions and commenting on the manuscript. Dr. Lan Yao (U. Michigan) is thanked for providing the code to compute the Kingham curves. The Cambridge Centre of Gallium Nitride is acknowledged as the source of the material used in some of these analyses, for work has been funded, in part, by the EPSRC (GR/S 49391/01) and these atom probe analyses were performed at the UK Atom-probe Facility at the University of Oxford — funded by the UK Engineering and Physical Sciences Research Council (EPSRC) under grant no. EP/077664/1.

## VIII. AUTHOR CONTRIBUTIONS

BG and DKS designed and coordinated the study, acquired and treated data, performed calculations and drafted the manuscript. DWS provided additional data and analysed GaN data, contributed to the discussion. MA and SS performed DFT calculations and discussed the results. ANC and MPM provided data, discussed the results and helped in drafting the manuscript. All authors read and commented on the manuscript.